\documentclass[prd,aps,showpacs,nofootinbib,preprint,eqsecnum]{revtex4}
\usepackage{graphicx,color,amsmath,amsxtra}
\usepackage{amssymb}
\usepackage[english]{babel}
\usepackage{amsfonts}

\def\e{{\rm e}}

\allowdisplaybreaks[4]
\begin{document}

\title{Cosmological perturbations in $F(R)$ gravity}

\author{Jiro Matsumoto$^{1}$\footnote{E-mail address: matumoto@th.phys.nagoya-u.ac.jp, jmatsumoto@tuhep.phys.tohoku.ac.jp}}
\affiliation{
$^1$Department of Physics, Nagoya University, Nagoya 464-8602, Japan\\
}

\begin{abstract}
The quasi-static solutions of the matter density perturbation in $F(R)$ gravity models have been 
investigated in numerous papers. 
However, the oscillating solutions in $F(R)$ gravity models have not been investigated enough so far. 
In this paper, the oscillating solutions are also examined by using appropriate approximations. 
And the behaviors of the matter density perturbation in $F(R)$ gravity models with singular evolutions of the physical parameters are 
shortly investigated as applications of the approximated calculations. 

\end{abstract}


\maketitle
\section{Introduction}

In order to explain the observational results about current accelerated expansion of the Universe, we need to modify 
the Einstein equation or introduce the anisotropic metric. The standard way to explain the accelerated expansion of the 
Universe is by introducing the new energy called ``dark energy," or revising the left-hand side of the Einstein equation to obtain ``modified gravity." 
The most popular model of these is the Lambda cold dark matter 
 ($\Lambda$CDM) model because of its simplicity. 
The $\Lambda$CDM model is given by introducing the contributions from the cosmological constant and the cold dark matter 
into the Einstein equation. 
However, an extraordinary fine-tuning for the value of the cosmological constant is necessary to explain the observational results, because 
the typical scale of gravity $M_{\mathrm{Pl}}$ is much larger than the observed scale of the critical density $\sim 10^{-3} \,$eV. 
Therefore, there have been suggested many dark energy models and modified gravity models to 
relax the fine-tuning problem. 
As an example of modified gravity models, there are $F(R)$ gravity models\cite{fr,Nojiri:2006ri,Sotiriou:2008rp,DeFelice:2010aj,Nojiri:2010wj}. 
$F(R)$ gravity models are described as an extension of the Einstein-Hilbert action to have higher power terms of the scalar curvature $R$. 
Today, viable $F(R)$ gravity models cannot be distinguished from the $\Lambda$CDM model by the observational data of the background evolution, 
so it is necessary to evaluate carefully the perturbations in these models. 
The perturbations in $F(R)$ gravity models had been also examined before the discovery of the current accelerated expansion of the Universe\cite{SS,Hwang}. 
However, we need to reconsider the perturbations 
because the matter can be negligible during the inflationary era of the early universe, but 
the matter is an essential component of the late-time Universe. 
And we are interested in the evolution of the matter density perturbation. 
In this paper, not only quasi-static solutions, but also oscillating solutions of the matter density perturbation are investigated 
by analytic calculations. 
By numerical calculations, 
this oscillating behavior of the matter density perturbation has been found in 
L.~Pogosian and A.~Silvestri\cite{oci1} and S.~Carloni,  E.~Elizalde, and S.~Odintsov\cite{oci2}. 
The behavior of the matter density perturbation in $F(R)$ gravity models with singular evolutions of the physical parameters are also investigated. 

In Sec.~II, the matter density perturbation is investigated by using some approximations and as applications, 
the perturbation is considered in the models which have singular behavior of some parameters. 
Section III begins from the derivation of the differential equation without using any approximations\cite{E}, and 
the behavior of the matter density perturbation is investigated while using the case analysis. 
The popular viable models of $F(R)$ gravity are considered in Sec.~IV. 
And concluding remarks are given in Sec.~V. 
The units of $k_\mathrm{B} = c = \hbar = 1$ are used and 
gravitational constant $8 \pi G$ is denoted by
${\kappa}^2 \equiv 8\pi/{M_{\mathrm{Pl}}}^2$ 
with the Planck mass of $M_{\mathrm{Pl}} = G^{-1/2} = 1.2 \times 10^{19}$GeV in this paper.
\section{Perturbative equations and approximated calculations}
\subsection{Background equations and linearized equations}
Here, we adopt the following action as a form of $F(R)$ gravity models: 
\begin{align}
S=\frac{1}{2 \kappa ^2} \int d^4 x \sqrt{-g} \left [ R+f(R) \right ]+S_\mathrm{matter}, 
\label{10}
\end{align}
where $f$ is an arbitrary function of the scalar curvature $R$, and $f(R)$ 
represents the deviation from the Einstein gravity. 
When we use the spatially flat Friedmann-Lemaitre-Robertson-Walker metric, $ds^2 = a^2(\eta) (d\eta^2-\sum_{i=1}^3 dx^i dx^i)$, 
the Friedmann-Lemaitre equations are written by 
\begin{align}
\frac{3\mathcal{H}'}{a^2}(1+f_{R})-\frac{1}{2}(R+f)
-\frac{3\mathcal{H}}{a^2}f_{R}' &=- \kappa ^2 \rho
\label{FL00}, \\
\frac{1}{a^2}(\mathcal{H}'+2\mathcal{H}^2)(1+f_{R})
-\frac{1}{2}(R+f)-\frac{1}{a^2}(\mathcal{H} f_{R}'+f_{R}'') &= \kappa^2 w \rho 
\label{FLii}, 
\end{align}
where $R = 6a^{-2} (\mathcal{H}'+\mathcal{H}^2)$, $f_{R} \equiv df(R)/dR$, and 
the prime represents the differentiation with respect to conformal time $\eta$. 
$\rho$ is the energy density of the matter coming from the variation of $S_\mathrm{matter}$ 
and $w$ is the equation of state parameter expressed by $w=p/ \rho$. 
The Hubble rate with respect to conformal time $\mathcal{H}$ is defined by $\mathcal{H}\equiv a'/a$. 
The equation of continuity is written as follows: 
\begin{equation}
\rho'+3(1+w)\mathcal{H}\rho = 0.   
\label{15}
\end{equation} 
The above equations (\ref{FL00})--(\ref{15}) are background equations of the Universe. 

On the other hand, $(00)$, $(ii)$, $(0i)=(i0)$, and $(ij)$, $i\neq j$ elements of the linearized Einstein equation in 
the Fourier space 
are represented as follows, respectively, when we use the metric of the Newtonian gauge, 
$ds^2 = a^2(\eta) [(1+2\Phi) d\eta^2- (1+2\Psi)\sum_{i=1}^3 dx^i dx^i ]$: 
\begin{align}
(1+f_{R}) \{ &-k^2(\Phi+\Psi)-3\mathcal{H}(\Phi'+\Psi')+(3\mathcal{H}'
-6\mathcal{H}^2)\Phi-3\mathcal{H}'\Psi \} \nonumber \\
&+f'_{R}(-9\mathcal{H}\Phi+3\mathcal{H}\Psi-3\Psi')\,=\,\kappa ^2 \rho a^2 \delta
\label{R00}, \\
(1+f_{R})\{ \Phi''+&\Psi''+3\mathcal{H}(\Phi'+\Psi')
+3\mathcal{H}'\Phi+(\mathcal{H}'+ 2\mathcal{H}^2)\Psi \} \nonumber \\
+&f'_{R}(3\mathcal{H}\Phi-\mathcal{H}\Psi+3\Phi') + f''_{R}(3\Phi-\Psi)\,=\,
c_{s}^{2} \kappa ^2 \rho a^2 \delta
\label{Rii}, \\
(1+f_{R})\{ \Phi'&+\Psi'+\mathcal{H}(\Phi+\Psi) \}+f'_{R}(2\Phi-\Psi)\,=\,
- \kappa ^2 \rho a^2 (1+w) v
\label{R0i}, \\
\Phi-\Psi - \frac{2f_{RR}}{a^2(1+f_{R})}& \{ 3\Psi''+6(\mathcal{H}'+\mathcal{H}^2)\Phi
+3\mathcal{H}(\Phi'+3\Psi')-k^2(\Phi-2\Psi) \} = 0
\label{Rij}.
\end{align}
Here, we define the sound speed by $c_\mathrm{s}^2 \equiv \delta p / \delta \rho$ and the matter density perturbation by 
$\delta \equiv \delta \rho / \rho$, 
to express Eqs.~(\ref{R00})--(\ref{Rij}). 
From the perturbation of the equation of continuity, $\nabla _\mu T^{\mu \nu}=0$, we have 
\begin{equation}
3\Psi'(1+w)-\delta'+3 \mathcal{H}(w-c_\mathrm{s}^2)\delta +k^2(1+w)v=0 
\label{T0} 
\end{equation}
and 
\begin{equation}
\Phi+\frac{c_{s}^2}{1+w}\delta+v'+\mathcal{H}v(1-3w)=0. 
\label{Ti}
\end{equation}
We treat the equation of state parameter and the sound speed of the matter as $w=c_\mathrm{s}=0$ by assuming 
the era after matter dominance. 

\subsection{Sub-horizon approximation} 
In the following, we derive the differential equation of the matter density perturbation by 
using sub-horizon approximation. Oscillating solutions cannot be, however, derived by such a procedure 
as shown in the $k$-essence model \cite{kperturbation}, so we will discuss them in the next subsection. 
Formally, sub-horizon approximation is given as $\partial/\partial \eta \sim \mathcal{H} \ll k$. 
It contains two approximations, which are small scale approximation $\mathcal{H} \ll k$ and quasi-static approximation 
$\partial/\partial \eta \sim \mathcal{H}$. 
From now on, we apply sub-horizon approximation to Eqs.~(\ref{R00})--(\ref{Rij}) to 
deduce the equation of the matter density perturbation; then it is necessary to be careful of the order of 
calculations. 
Unlike in the case of the exact calculations, the procedure to derive the equation of the matter density 
perturbation is not unique in the approximated calculations. 
For example, we can choose at least two procedures to derive the equation with respect to $\delta$ when 
$\vert f_\mathrm{RR} \vert k^2/a^2 \ll 1$. 
The first procedure is as follows: 
\begin{enumerate}
 \item Combining Eqs.~(\ref{T0}) and (\ref{Ti}) to erase $v$. 
 \item From Eq.~(\ref{Rij}), reading a relation $\Phi \thickapprox \Psi$. 
 \item By using a sub-horizon approximation, we obtain $-2 k^2(1+f_R)\Phi \thickapprox 
 \kappa ^2 \rho a^2 \delta$ from Eq.~(\ref{R00}). 
 \item Gathering the above equations gives the equation only described by the parameter $\delta$.  
\end{enumerate}
The other procedure is as follows:
\begin{enumerate}
 \item Combining Eqs.~(\ref{R0i}) and (\ref{T0}) to erase $v$.
 \item From Eq.~(\ref{Rij}), reading a relation $\Phi \thickapprox \Psi$. 
 \item By using a sub-horizon approximation, we obtain $-2 k^2(1+f_R)\Phi \thickapprox 
 \kappa ^2 \rho a^2 \delta$ from Eq.~(\ref{R00}). 
 \item Gathering the above equations gives the equation only described by the parameter $\delta$.  
\end{enumerate} 
The only difference between the two procedures is the first step. 
Because of it, the derived equations, however, become quite different. 
So we should be sure of which is the correct equation. The answer is obvious. 
The approximated calculations we considered by now are only correct at the leading order, but the cancellation of 
the leading order happened in the second procedure, so that the correct equation is the first one: 
\begin{align}
\delta '' + \mathcal{H} \delta ' - \frac{3a^2 \Omega_{\mathrm m}}{2(1+f_R)} \delta = 0. 
\label{301}
\end{align}
Similarly, when $\vert f_\mathrm{RR} \vert k^2/a^2 \gg 1$, we have the following equation by using sub-horizon 
approximation: 
\begin{align}
\delta '' + \mathcal{H} \delta ' - \frac{2a^2 \Omega_{\mathrm m}}{(1+f_R)} \delta = 0, 
\label{302}
\end{align}
because a relation $\Phi \thickapprox 2 \Psi$ can be read from Eq.(\ref{Rij}). 

\subsection{The approximation of ultra-relativistic variation}
As referred in the last subsection, oscillating solutions cannot be derived by sub-horizon approximation, 
so we use the approximation of ultra-relativistic variation $\partial/\partial \eta \sim k$ and 
small scale approximation $\mathcal{H} \ll k$ instead. 
To obtain the equation of $\delta$, we first assume a relation $\Phi \thickapprox -\Psi$ 
from the leading terms of Eq. (\ref{Rii}). And then, Eq. (\ref{Rij}) gives 
\begin{align}
\Psi + \frac{3 f_{RR}}{a^2(1+f_R)}(\Psi'' + k^2 \Psi) \thickapprox 0. 
\label{303}
\end{align}
If $\vert f_{RR} \vert k^2/a^2 \ll 1$, there is an obvious solution so that 
the $F(R)$ gravity models which satisfy the condition $\vert f_{RR} \vert k^2/a^2 \ll 1$ do not have 
oscillating solutions of $\Psi$ and $\delta$ within the form of $\mathrm{e}^{ik\eta}$. 
On the other hand, if $\vert f_{RR} \vert k^2/a^2 \gg 1$, $\Psi$ should satisfy the equation, 
$\Psi'' + k^2 \Psi \thickapprox 0$. 
The solutions of it is, 
\begin{align}
\Psi = C_1 \mathrm{e}^{-ik \eta} + C_2 \mathrm{e}^{ik \eta}, 
\label{304}
\end{align} 
where $C_1$ and $C_2$ are integration constants. 
Then, combining Eqs. (\ref{T0}) and (\ref{Ti}) gives a relation between $\delta$ and $\Psi$ as 
$\delta '' \thickapprox 2 \Psi ''$, so we have solutions of $\delta$: 
\begin{align}
\delta = 2C_1 \mathrm{e}^{-ik \eta} + 2C_2 \mathrm{e}^{ik \eta} + C_3 \eta + C_4, 
\label{305}
\end{align}
where $C_3$ and $C_4$ are integration constants. 
However, the behavior of the oscillating solutions are not only determined by the 
leading order terms. Therefore, calculations without approximations are necessary for correct evaluations. 
\subsection{The behavior near the singularity} 
It is known that there are several future singularity scenarios of the Universe\cite{singular, reconstruction}, 
\begin{itemize}
\item Type I (''Big Rip'') : For $t \to t_s$, $a \to \infty$,
$\rho_\mathrm{eff} \to \infty$, and $\left|p_\mathrm{eff}\right| \to \infty$.
This also includes the case of $\rho_\mathrm{eff}$, $p_\mathrm{eff}$ being
finite at $t_s$.
\item Type II (''Sudden'') : For $t \to t_s$, $a \to a_s$,
$\rho_\mathrm{eff} \to \rho_s$, and $\left|p_\mathrm{eff}\right| \to \infty$. 
\item Type III : For $t \to t_s$, $a \to a_s$,
$\rho_\mathrm{eff} \to \infty$, and $\left|p_\mathrm{eff}\right| \to \infty$. 
\item Type IV : For $t \to t_s$, $a \to a_s$,
$\rho_\mathrm{eff} \to 0$, $\left|p_\mathrm{eff}\right| \to 0$, and higher
derivatives of $H$ diverge.
This also includes the case in which $p_\mathrm{eff}$ ($\rho_\mathrm{eff}$)
or both of $p_\mathrm{eff}$ and $\rho_\mathrm{eff}$
tend to some finite values, while higher derivatives of $H$ diverge.
\end{itemize}
And lately the little rip\cite{little} scenario is discussed, where any finite singularity does not appear but causes 
the destruction of the bound system. 
In the following, we will investigate the behavior of the quasi-static solutions of the matter density perturbation near these singularities. 
However, we do not consider Type I, Type III, and the little rip scenarios, because the approximations in 
these scenarios are equivalent to the superhorizon approximation. 
\subsubsection*{Type II}
In the case of Type II, the Hubble rate $H \equiv \dot a(t)/ a(t)$ is finite, but the derivative of the Hubble rate $\dot H$ is singular, so 
the approximations $k^2$, $\mathcal{H}^2 \ll \vert \mathcal{H}' \vert$ are applied in the following. 
And the quasi-static approximation $\partial _\eta \sim \mathcal{H}$ will be used. 
We first consider Eq.~(\ref{Rij}) to obtain the equation with respect to $\delta$, because the equation $\Psi '' \thickapprox -2 \mathcal{H}' \Phi$ can be read 
by considering relations $\Psi '' \sim (\mathcal{H} \Psi)' \sim \mathcal{H}' \Psi$. 
And $2\Phi \thickapprox \Psi$ is assumed from Eqs.~(\ref{R0i}) and (\ref{T0}). Then, the equation of $\Psi$: $\Psi '' + \mathcal{H}' \Psi = 0$ is obtained. 
The solutions of the equation can be written as $\Psi = const./a +const. \mathrm{e}^{-a}$, then $\Psi ' = - \mathcal{H} \Psi$. 
By using Eq.~(\ref{R00}) and $\Psi ' = - \mathcal{H} \Psi$, we have the following equation: 
\begin{align}
\Psi'' \thickapprox \frac{2 \mathcal{H}'}{3 \mathcal{H}' - 7 \mathcal{H} f_R'} \frac{\kappa ^2 \rho a^2}{1+f_R} \delta. 
\label{IIa}
\end{align}
Here, the assumption $2\Phi \thickapprox \Psi$ is justified because Eq.~(\ref{IIa}) implies that 
$\Psi \gg \vert \mathcal{H}^4 \delta/(k^2 \mathcal{H}') \vert$. 
To erase the parameter $v$ from Eqs.~(\ref{T0}) and (\ref{Ti}) gives the following equation: 
\begin{align}
\delta '' + \mathcal{H} \delta ' -3 \Psi '' -3 \mathcal{H} \Psi ' + k^2 \Phi = 0, 
\label{IIb}
\end{align}
so that the substitution of Eq.~(\ref{IIa}) into Eq.~(\ref{IIb}) gives the equation of the matter density perturbation, 
\begin{align}
\delta '' + \mathcal{H} \delta ' - \frac{6 \mathcal{H}'}{3 \mathcal{H}' - 7 \mathcal{H} f_R'} \frac{\kappa ^2 \rho a^2}{1+f_R} \delta \thickapprox 0. 
\label{IIc}
\end{align}
Equation (\ref{IIc}) indicates that the behavior of the matter density perturbation changes near the singularity 
because Eq.~(\ref{IIc}) is different from Eqs.~(\ref{301}) and (\ref{302}). 
\subsubsection*{Type IV} 
In case of type IV, the approximations $k^3$, $\mathcal{H}^3$, $\mathcal{H} \vert \mathcal{H}' \vert \ll \vert \mathcal{H}'' \vert$ should be applied, and 
from Eqs.~(\ref{FL00}) and (\ref{FLii}) $\mathcal{H}^2 \vert f_{RR} \vert$, $\mathcal{H}^4 \vert f_{RRR} \vert \ll 1$ are satisfied. 
Then, Eq.~(\ref{Rij}) gives $\Phi \thickapprox \Psi$. Equations (\ref{R00}), (\ref{R0i}), and (\ref{T0}) give 
\begin{align}
\Psi \thickapprox \Phi \thickapprox \frac{\kappa ^2 \rho a^2}{1+f_R} \left \{ -4k^2 
+ \frac{6 \kappa ^2 \rho a^2}{1+f_R} + \frac{18 \kappa ^2 \rho \mathcal{H}^2 a^2}{(1+f_R)k^2}  \right \} ^{-1} 
\left [ \left \{ 2- \frac{3 \kappa ^2 \rho a^2}{(1+f_R)k^2} \right \} \delta - \frac{6 \mathcal{H}}{k^2} \delta ' \right ]. 
\label{IVa} 
\end{align} 
The substitution of Eq.~(\ref{IVa}) into  Eq.~(\ref{IIb}) will give the third order differential equation with respect to $\delta$, but 
it is not explicitly expressed here. 

There is a notable literature written by A.~de~la~Cruz-Dombriz \textit{et al.} \cite{E}, 
which expresses a strict derivation of the differential equation of the matter density 
perturbation in $F(R)$ gravity models. 
However, it is too difficult to analyze the time evolution of the matter density perturbation exactly 
because the differential equation is so complicated. 
We use some approximations in the following. We then need to be careful to the order of applying approximations, 
because their order is not commutable. 

\section{Analyses started from the fourth order equation}
\subsection*{Case I}
First, the differential equation of the matter density perturbation 
$\delta \rho/ \rho$ when we use small scale approximation $\mathcal{H}^2 \ll k^2$ 
is derived by using Eq.~(31) and the Appendix of \cite{E} as follows: 
\begin{align}
\delta '''' &+ \mathcal{H} \left ( 3+\frac{f_R'}{\mathcal{H}(1+f_R)} +O(\mathcal{H}^2/k^2) \right )\delta '''  
+\mathcal{H}^2 \left ( \frac{k^2}{\mathcal{H}^2} +  O(1) \right ) \delta '' \nonumber \\
&+ \mathcal{H}^3 \left ( \frac{k^2}{ \mathcal{H}^2} +  O(1) \right ) \delta ' 
-\mathcal{H}^4 \left ( \frac{2 \kappa^2 k^2 \rho a^2}{3\mathcal{H}^4(1+f_R)} + O(1) \right )\delta =0, 
\label{20}
\end{align} 
where we have neglected sub-leading terms under the approximation $\mathcal{H}^2 \ll k^2$. 
If we assume quasi-static solutions of $\delta$, then we obtain an equation from the terms proportional to $k^2$ in Eq.~(\ref{20}): 
\begin{equation}
\frac{d^2 \delta}{dN^2} + \left ( \frac{1}{2} - \frac{3}{2}w_\mathrm{eff} \right ) \frac{d \delta}{dN}
 -\frac{2}{1+f_R} \Omega_\mathrm{m} \delta =0. 
\label{30} 
\end{equation}
Here, we have defined $w_\mathrm{eff}= -2 \dot H/(3H^2) -1$, $\Omega_m = \kappa ^2 \rho /(3 H^2)$, and $N \equiv \ln a$. 
And $H$ means the standard Hubble rate $H \equiv \dot a(t)/ a(t)$. 
In the matter dominant era, Eq.~(\ref{30}) becomes 
\begin{equation}
\frac{d^2 \delta}{dN^2} + \frac{1}{2} \frac{d \delta}{dN} -\frac{2}{1+f_R}  \delta =0. 
\label{40}
\end{equation}
The growth factor of $\delta$, $f\equiv d\ln \vert \delta \vert /dN$, 
is given by $f=-1/4 \pm \sqrt{33}/4$ when $\vert f_R \vert \ll 1$ from Eq.~(\ref{40}). 
Therefore, the growth factor of this case in the matter dominant era is bigger than 
that of $\Lambda$CDM model, $f_\mathrm{max} = 1$. 
On the other hand, the fourth order differential equation (\ref{20}) has the following oscillating solutions: 
\begin{align}
\delta (\eta) &= C_1(\eta){\rm e}^{ik \int d \eta} 
+ C_2(\eta) {\rm e}^{-ik \int d \eta} \nonumber \\
&= const. \times \frac{1}{a \sqrt{1+f_R}}{\rm e}^{ik \int d \eta} 
+ const. \times \frac{1}{a \sqrt{1+f_R}}{\rm e}^{-ik \int d \eta} \label{50}
\end{align}
We have used the WKB approximation under the condition $\mathcal{H}^2 \ll k^2$. 
It is useful to define the effective growth factor 
$f_\mathrm{eff}$ as $f_\mathrm{eff} \equiv d\ln \vert C_1 \vert/dN = d\ln \vert C_2 \vert/N$
because the effective time evolution of the solution (\ref{50}) is determined by $C_1(\eta)$ or $C_2(\eta)$. 
In this case, the effective growth factor is given by 
\begin{equation}
f_\mathrm{eff}=-1-3\left ( \frac{\ddot H}{H^3} + 4\frac{\dot H}{H^2} \right )\frac{H^2 f_{RR}}{1+f_R}, 
\label{60}
\end{equation}
which depends on the property of each model. If we consider some models of $F(R)$ gravity, which 
does not satisfy the condition $\vert H^2 f_{RR}\vert \ll 1$, the oscillating solutions can 
be growing ones. 

\subsection*{Case II}
Next, we consider the time evolution of the matter density perturbation in the case of 
$(k/\mathcal{H})^2\vert f_\mathrm{R} \vert \ll 1$, $\vert f_\mathrm{RR} \vert k^2/a^2 \gg 1$ and $\mathcal{H}^2 \ll k^2$. 
We can also obtain the following equation of matter density perturbation in this case 
by using Eq.~(31) and the Appendix in \cite{E}: 
\begin{align}
\delta '''' &+  \mathcal{H} \left ( 3+O(\mathcal{H}^2/k^2) \right ) \delta ''' 
+\mathcal{H} ^2 \left ( k^2/\mathcal{H}^2 +O(1)  \right ) \delta '' \nonumber \\
&+ \mathcal{H}^3 \left ( k^2/\mathcal{H}^2 +O(1) \right ) \delta ' 
- \mathcal{H}^4 \left \{ \frac{4k^2}{3 \mathcal{H}^2} \left ( 1- \frac{\mathcal{H}'}{\mathcal{H}^2} \right )
+O(1) \right \} \delta  =0. 
\label{70}
\end{align}
If we note to the terms proportional to $k^2$, then we have 
\begin{equation}
\frac{d^2 \delta}{dN^2} + \left ( \frac{1}{2} - \frac{3}{2}w_\mathrm{eff} \right ) \frac{d \delta}{dN}
 -2\left ( 1+w_\mathrm{eff} \right ) \delta =0. 
\label{80}
\end{equation}
In the matter dominant era, the equation 
\begin{equation}
\frac{d^2 \delta}{dN^2} + \frac{1}{2} \frac{d \delta}{dN} -2 \delta =0 
\label{90}
\end{equation}
holds. The growth factors of this quasi-static solutions are $f=-1/4\pm \sqrt{33}/4$, which are same as 
those in Case I. 
While, the oscillating solutions are given as 
\begin{equation}
\delta (\eta) = C_1 \frac{1}{a }{\rm e}^{ik \int d \eta} + C_2 \frac{1}{a }{\rm e}^{-ik \int d \eta}, 
\label{100}
\end{equation}
where $C_1$ and $C_2$ are arbitrary constants. 
The effective growth factor is expressed by
\begin{equation}
f_\mathrm{eff}=-1. 
\label{110}
\end{equation}
If $\vert H^2 f_{RR}\vert \ll 1$, Eq.~(\ref{110}) is same as Eq.~(\ref{60}), while if 
$H^2 f_{RR}$ is not negligible, it is different from that in Case I. 

\subsection*{Case III}
If $(k/\mathcal{H})^2\vert f_\mathrm{R} \vert \ll 1$, $\vert f_\mathrm{RR} \vert k^2/a^2 \ll 1$ and $\mathcal{H}^2 \ll k^2$, 
the fourth order differential equation of $\delta$ is given by 
\begin{align}
\delta '''' &+ \left \{ \frac{ 12\mathcal{H}^2 (-2+ \mathcal{H}''/\mathcal{H}^3)f_{RRR}}{a^2f_{RR}}
+\frac{1-\mathcal{H}'/\mathcal{H}^2}{-2+ \mathcal{H}''/\mathcal{H}^3} + O(\mathcal{H}^2/\chi^2) \right \} \mathcal{H}
 \delta ''' \nonumber \\
+& \chi^2 \left \{
\left ( 1 + O(\mathcal{H}^2/\chi^2) \right ) \delta '' 
+ \mathcal{H} \left ( 1 + O(\mathcal{H}^2/\chi^2) \right )  \delta ' 
+ \mathcal{H}^2  \left ( 2 \frac{\mathcal{H}'}{\mathcal{H}^2}-  \frac{\mathcal{H}''}{\mathcal{H}^3} 
 + O(\mathcal{H}^2/\chi^2) \right ) \delta  \right \} =0 
\label{fr120}, \\
& \chi \equiv \sqrt{\frac{a^2}{3f_{RR}}\frac{1-\mathcal{H}'/\mathcal{H}^2}{2- \mathcal{H}''/\mathcal{H}^3}}.
\label{fr130}
\end{align}
Noting to the terms proportional to $\chi^2$, we obtain 
\begin{equation}
\frac{d^2 \delta}{dN^2} + \left ( \frac{1}{2} - \frac{3}{2}w_\mathrm{eff} \right ) \frac{d \delta}{dN}
 +\left ( 2\frac{\dot H}{H^2}+\frac{\ddot H}{H^3} \right ) \delta =0. 
\label{fr140}
\end{equation}
The solutions of Eq.~(\ref{fr140}) depend not only on $f_R$ and $f_{RR}$ but also on 
$f_{RRR}$ and $f_{RRRR}$, because Friedmann-Lemaitre equations (\ref{FL00}) and (\ref{FLii}) yield the 
following relation: 
\begin{equation}
 2\frac{\dot H}{H^2}+\frac{\ddot H}{H^3} = -\frac{3 \Omega_\mathrm{m}}{2(1+f_R)} - \frac{f_R'}{\mathcal{H}(1+f_R)}
 - \frac{f_R''}{\mathcal{H}^2(1+f_R)}+ \frac{f_R'''}{2\mathcal{H}^3(1+f_R)}. 
\label{fr150}
\end{equation}
On the other hand, the oscillating solutions are given as: 
\begin{equation}
\delta (\eta) = C_1 {\rm e}^{\int f_\mathrm{eff} dN +i \int \chi d \eta} 
+ C_2 {\rm e}^{\int f_\mathrm{eff} dN -i \int \chi d \eta}. 
\label{fr160}
\end{equation}
Here, $C_1$ and $C_2$ are arbitrary constants, and the effective growth factor $f_\mathrm{eff}$ 
is defined as: 
\begin{equation}
f_\mathrm{eff}=1-\frac{5}{2} \frac{d}{dN} \ln \vert \chi \vert -2 \frac{d}{dN} \ln \vert f_{RR} \vert
 + \frac{1-\mathcal{H}'/\mathcal{H}^2}{2-\mathcal{H}''/\mathcal{H}^3}. 
\label{fr170}
\end{equation}
We cannot further discuss the behavior of the solutions without assuming some models, because 
they also depend on higher derivatives of $f(R)$. 

\subsection{Model dependent analysis}
\subsubsection{Hu and Sawicki's model}
We consider here the following form of the function $f(R)$: 
\begin{align} 
f(R) = - m^2 \frac{c_1 (R/m^2)^n}{c_2 (R/m^2)^n +1}, 
\label{HS}
\end{align}
which was proposed by W.~Hu and I.~Sawicki \cite{H&S}. 
If we use the best fit model\cite{N} to the Sloan Digital Sky Survey (SDSS) and the seven year data of the
 Wilkinson Microwave Anisotropy Probe (WMAP7), we can examine detailed time dependence of the matter density 
perturbation. In this case, parameters of the model are given by $n=1.53$, $c_1=10^{3.47}$, and $c_2=10^{2.28}$, 
respectively, then this model is classified to Case III by assuming the scale, $k \sim 0.1 - 0.01 h\, \mathrm{Mpc}^{-1}$. 
Therefore, the deviation from the $\Lambda$CDM model is a little at the perturbation order when we consider the quasi-static mode because 
$R^3f_{RRR}$ and $R^4f_{RRRR}$ are also negligible in these parameter sets. 
This result is consistent with the conclusions in \cite{N}. 
The fast fluctuating mode will be considered a bit later. 
\subsubsection{Starobinsky's model} 
We next consider the Starobinsky's model \cite{Starobinsky} 
\begin{align}
f(R) = \lambda R_0 \left ( \left ( 1+\frac{R^2}{R_0^2} \right )^{-n} -1 \right ). 
\label{Sta}
\end{align}
By the same procedure, we obtain the same result with Hu and Sawicki's model, i.e. the deviation from the 
$\Lambda$CDM model is so small that the observational verification is difficult when we consider the quasi-static mode. 

Such a result comes from the necessity to satisfy the local gravity tests. 
Namely, fitting the models of $F(R)$ gravity to the general relativity in the solar system causes 
the deviation from the $\Lambda$CDM model a little also in the large scale. 
It is easy to understand where this result comes from by referring to Sec.~IIB. 
That is to say, the condition $\vert f_\mathrm{RR} \vert k^2/a^2 \ll 1$ induces this result. 
\subsubsection{Exponential gravity model}
Next we consider the exponential type function of $f(R)$\cite{Exp1,Exp2,Exp3}: 
\begin{align}
f(R) = -cr(1-\e ^{-R/r}), 
\label{exp}
\end{align}
which has a possibility to deviate from the $\Lambda$CDM model, because of its sharp dependence of $R$. 
Some studies constrain the parameters in Eq.~(\ref{exp})\cite{AGSS,YLLG}; however, there is no 
effective constraint on large $c$. 
If we consider $c \sim 1000$, we would find interesting behavior of $\delta$ of the quasi-static mode which comes from 
$\vert R^3f_{RRR} \vert \sim 1$
, while $\vert Rf_R \vert, \vert R^2f_{RR} \vert \ll 1$. 
However, it is difficult to cause the characteristic evolution of the matter density perturbation without changing the 
background evolution of the Universe, because $\vert R^3f_{RRR} \vert \sim 1$ induces the changes in the background equations 
(\ref{FL00}) and (\ref{FLii}).

\subsubsection{The fast fluctuating mode} 
We have seen that the $F(R)$ gravity models which yield almost equivalent background evolution to that of the $\Lambda$CDM model 
cannot be also distinguished from the $\Lambda$CDM model at perturbation level when we consider 
the quasi-static mode of the matter density perturbation. 
So the other mode of the matter density perturbation, the fast fluctuating mode, is investigated in the following. 
The models which yield almost equivalent background evolution to that of the $\Lambda$CDM model 
are classified as Case III, i.e. the case of $(k/\mathcal{H})^2\vert f_\mathrm{R} \vert \ll 1$, 
$\vert f_\mathrm{RR} \vert k^2/a^2 \ll 1$, and $\mathcal{H}^2 \ll k^2$. 
It is determined by the argument of $\chi$ whether or not the fast fluctuating solutions are oscillating solutions, 
because the behavior of the fast fluctuating solutions are understood from Eqs.~(\ref{fr160}) and (\ref{fr170}), 
The argument of $\chi$ is found from the sign of $f_{RR}$ 
because the relation, $(1-\mathcal{H}'/\mathcal{H}^2)/(2- \mathcal{H}''/\mathcal{H}^3) \sim 1$, is held in this case. 
On the other hand, the condition $f_{RR}>0$ is imposed to the famous viable $F(R)$ models to 
satisfy the scalaron\cite{scalaron} mass squared to be positive. Namely, $\chi$ is real. 
Therefore, the fast fluctuating solutions always express oscillating behavior in those models. 
Next, it is important to investigate whether or not this oscillation is decaying oscillation. 
The expression of the effective growth factor (\ref{fr170}) enables us to know the behavior of the oscillation. 
By using Eqs.~(\ref{FL00}) and (\ref{FLii}) we have 
\begin{align}
f_\mathrm{eff} \simeq -\frac{1}{2} + \frac{9}{2} \left ( 2-\frac{\mathcal{H}''}{\mathcal{H}^3} \right ) \frac{\mathcal{H}^2 f_{RRR}}{a^2f_{RR}}. 
\label{fre1}
\end{align}
Here, $\mathcal{H}''/\mathcal{H}^3\simeq 1/2$ is held in the matter dominant era. 
Thus, it is determined by $f_{RR}$ and $f_{RRR}$ that whether or not $f_{\mathrm{eff}}$ is positive. 
If the values of parameters in \cite{N} are applied to Hu and Sawicki's model and Starobinsky's model, 
we obtain $f_{\mathrm{eff}}<0$ because of $f_{RR}/f_{RRR} < 0$, 
i.e. it indicates these solutions are decaying oscillating solutions. 
This result does not strongly depend on the value of the parameters because the condition $f_{RR}/f_{RRR}<0$ is 
derived from the property that $f_R$ is approximately regarded as an inverse power law function of $R$ in these models. 
At the same time, the oscillating solutions are also decaying solutions in the exponential gravity model 
because $f_{RR}/f_{RRR}<0$ is satisfied without depending on the values of parameters. 

The period of this oscillation is shorter than that of quintessence model\cite{kperturbation}, as we can understand from 
the condition $\vert f_\mathrm{RR} \vert k^2/a^2 \ll 1$. 
That is to say, this oscillation corresponds to the oscillation when the sound speed exceeds light speed in the $k$-essence model.
However, this oscillation does not mean superluminality in $F(R)$ gravity. Therefore, there are not problems about causality. 

How to yield these solutions can be supposed from Eq.~(\ref{Rij}). 
The ($ij$), $i\neq j$ element of the linearized Einstein equation simply yields $\Phi = \Psi$ in the $\Lambda$CDM model and $k$-essence model. 
However, there are solutions, satisfying $\Phi = \Psi$ or $\Psi \sim f_{RR} \Psi ''/a^2$ in $F(R)$ gravity models. 
And the latter solutions correspond to the fast fluctuating mode. 

\section{Conclusions}
First, the time evolution of the matter density perturbation has been investigated by using approximations. 
When we use the sub-horizon approximation, it has been shown that the evolutionary behavior of the matter density perturbation depends on 
which condition is satisfied $\vert f_\mathrm{RR} \vert k^2/a^2 \ll 1$ or $\vert f_\mathrm{RR} \vert k^2/a^2 \gg 1$, as is well known. 
On the other hand, we may obtain the oscillating solutions by using the approximation $\partial/\partial \eta \sim k$ instead of 
the sub-horizon approximation. 
And the variation of the matter density perturbation near the singularity has also been investigated 
as expanded applications of the approximate calculations. 

Next, the growth of the matter density perturbations has been investigated by using the results in \cite{E} and 
the case analysis.
The first case is that the small scale approximation, $\mathcal{H}^2 \ll k^2$, is more valid than others. 
The second case is that the approximation $\vert f_\mathrm{R} \vert$ is small enough, and is more valid than small scale approximation, but 
$\vert \mathcal{H}^2 f_\mathrm{RR} \vert$ is not very small. 
The third case is that approximations $\vert f_\mathrm{R} \vert \ll 1$ and $\vert \mathcal{H}^2 f_\mathrm{RR} \vert \ll 1$ 
are more valid than the small scale approximation. 
In each of the cases, we have obtained characteristic solutions. 
Especially, the following behaviors have been shown: the quasi-static solutions in the first case grow faster than that of the $\Lambda$CDM model, and 
the fast fluctuating solutions in the third case have nontrivial dependence of the function $f(R)$.
The popular viable models correspond to the third case if the parameters in these models are determined to 
fit to the observations of the background evolution of the Universe. 
And the quasi-static solutions are equivalent to that of the $\Lambda$CDM model. 
Moreover, the fast fluctuating solutions in those models are the decaying oscillating solutions. 
Thus, it is not possible to constrain these models from the matter density perturbation if 
the $\Lambda$CDM model indicates the true evolution of the background universe and the matter density perturbation. 

The conditions for the fast fluctuating mode to be decaying oscillating solutions are that of $f_{RR}>0$ and $f_{RRR}<0$ for the models
, which yield almost equivalent background evolution to the $\Lambda$CDM model. 
So it can be said that imposing $f_{RR}/f_{RRR}>0$ enables us to make the model, which yields almost equivalent background evolution 
to the $\Lambda$CDM model, but yields different evolution of the matter density perturbation compared to the $\Lambda$CDM model. 


\section*{Acknowledgments}
The author is grateful to S.~Nojiri for corrections and advice, and also thanks S.D.~Odintsov for some comments. 
This work was supported in part by the Global COE Program of Nagoya University, 
''Quest for Fundamental Principles in the Universe (QFPU)'' from JSPS and MEXT of Japan 
and supported by Grant-in-Aid for JSPS Fellows \#461001209. 


\end{document}